\documentclass{article}
\usepackage{graphicx}
\usepackage{amsmath}
\usepackage{amsfonts}
\usepackage{amssymb}
\newtheorem{theorem}{Theorem}

\newtheorem{conjecture}[theorem]{Conjecture}

\begin{document}

\title{Thermal behavior induced by vacuum polarization on causal horizons in
comparison with the standard heat bath formalism\footnote{To appear in the
proceedings of the ''6th International Workshop on Conformal Field Theory and
Integrable Models'', in Chernogolovka, September 2002}\\{\small Dedicated to A. A. Belavin on the occasion of his 60th birthday}}
\author{Bert Schroer\\present address: CBPF, Rua Dr. Xavier Sigaud 150, \\22290-180 Rio de Janeiro, Brazil\\email schroer@cbpf.br\\permanent address: Institut f\"{u}r Theoretische Physik\\FU-Berlin, Arnimallee 14, 14195 Berlin, Germany}
\date{Jannuray 2003}
\maketitle
\begin{abstract}
Modular theory of operator algebras and the associated KMS property are used
to obtain a unified description for the thermal aspects of the standard heat
bath situation and those caused by quantum vacuum fluctuations from
localization. An algebraic variant of lightfront holography reveals that the
vacuum polarization on wedge horizons is compressed into the lightray
direction. Their absence in the transverse direction is the prerequisite to an
area (generalized Bekenstein-) behavior of entropy-like measures which reveal
the loss of purity due to restrictions to wedges and their horizons. Besides
the well-known fact that localization-induced (generalized Hawking-)
temperature is fixed by the geometric aspects, this area behavior (versus the
standard volume dependence) constitutes the main difference between
localization-caused and standard thermal behavior.
\end{abstract}

\section{Introductory remarks}

In contradistinction to the standard Boltzmann-Gibbs stochastic
framework\footnote{For brevity the standard thermal setting will be referred
to as the ``heat bath'' formalism, as opposed to thermal manifestations of
\ ''localization'' whose explanation and investigation constitutes the main
topic of this lecture.} of ensembles coupled to a heat bath, which originated
in the setting of classical physics, the recognition that QFT via its
characteristic vacuum polarization effects caused by localization may lead to
purely quantum thermal manifestations (even if the global state is a ground
state i.e. not a member of an ensemble) is a more recent theoretical discovery
which started with Hawking's observation on black hole radiation
\cite{Hawking}. Although this observation in its original form was dependent
on some classical geometrical prerequisites in particular on nontrivial
spacetime curvature leading to a Schwarzschild-like black hole of finite
radius, it became gradually clear that this thermal manifestation was really a
characteristic property for local quantum physics since it is potentially
present in all quantum theories (different from quantum mechanics) with a
built-in causal propagation structure as a result of a maximal propagation
velocity. It is not even limited to the kind of field theoretic models used in
relativistic particle physics, but extends to black hole'' analog quantum
systems \cite{U} in acoustics (phonon excitations), quantum optics or
hydrodynamics; in short to all quantum system which admit a Fock space
structure with causally propagating quantum excitations. Although the
mathematical manifestation of thermal properties in such a local quantum
physics setting is omnipresent (see the Tomita-Takesaki theorem in conjunction
with the quantum field theoretic Reeh-Schlieder property in the next section),
it is not so easy to find conditions under which it leads to observable
physical consequences since (at least on the present level of understanding)
physical manifestations require very special geometric circumstances. A prime
example, which also will serve as our main illustration, is the Gedanken
experiment of Unruh \cite{Unruh} in which the causal horizon of a Rindler
wedge region in Minkowski spacetime is realized with the help of a family of
uniformly accelerated observers (whose worldlines stay inside a wedge such
that the (upper) horizon of the wedge defines a ``causal protection'' against
any backreaction of the outside world). In Minkowski spacetime this wedge
situation is the only region which admits a geometrical symmetry (a
diffeomorphism associated with a Killing vector) which induces an automorphism
of the algebra of observables which is generated by local fields associated
with the chosen localization region (in conformal invariant theories the
modular group of double cone algebras is also geometric \cite{Haag})

At the time of Hawking's and Unruh's discovery there was an unrelated general
structural observation within QFT made by Bisognano and Wichmann \cite{Haag}.
Its main point consisted in demonstrating that the application of the
Tomita-Takesaki modular theory \cite{Bra-Ro}\cite{Borchers} to the operator
algebra $\mathcal{A}(W)$ generated by fields smeared with test functions with
support in the wedge leads to a physical identification of the modular objects
associated with the pair ($\mathcal{A}(W),\Omega)$ with $\Omega$ being the
vacuum state vector. The modular group turned out to be the wedge associated
Lorentz boost and the Tomita involution became identified (up to spatial
rotation) with the all important TCP operation. The T-T modular theory was a
significant mathematical conquest made one decade before, whose physical
significance for the conceptual basis of infinite volume quantum statistical
mechanics was a completely independent seminal contribution of Haag, Hugenholz
and Winnink\footnote{One of the important observations in the HHW paper was
that the Kubo-Martin-Schwinger condition (which was proposed as an analytic
trick to avoid the computation of traces) played a fundamental conceptual role
in the formulation of statistical mechanics of open systems and (as was later
shown in based on observations in) for the notion of stability of equilibrium
states \cite{Haag}.}. The new aspect of the Bisognano-Wichmann observation was
that the KMS property, which, as the H-H-W paper had shown, was characteristic
for equilibrium thermal quantum physics in the thermodynamic limit (i.e. when
the Gibbs state formulation looses its mathematical meaning), does become (via
modular theory) also an attribute of the wedge-restricted vacuum state.
Different from the standard heat bath Gibbs ensemble formalism, the
temperature of the restricted vacuum state (which thermalizes through its
restriction from the global algebra to the wedge localized subalgebra) is not
a free parameter but, as in Hawking's black hole case or in Unruh's
illustration, is determined by the geometry of localization. It did not take
long before the deep connection of the Hawking-Unruh aspect with the KMS
property of localization as observed by Bisognano and Wichmann was made
\cite{Sewell}; this was followed by a steady stream of papers in which the
relation of thermal aspects of quantum matter with their geometrical
localization behind causal horizons\footnote{The event horizons related to
Killing vectors in GR are causal horizons, but the latter (which arise as
boundaries after causal completion of arbitrary spacetime regions) are not
restricted to Killing symmetries (example: double cones in Minkowski
spacetime).} was investigated (see the extensive list of reference in
\cite{Borchers}.

The aim of these notes is to review the role of the KMS property and modular
theory as the unifying formalism behind both the conventional heat bath- and
the more subtle localization- induced thermal properties, for an extensive
article with a similar agenda see \cite{Fröhlich}. Recent studies
\cite{Schroer} have shown that further insight into thermal aspects of
localized matter can be obtained by using an algebraic adaptation of the idea
of holographic\cite{Hooft} projection. The holographic projection of a wedge
algebra consists of an algebra on its (upper) lightfront horizon (half of the
lightfront which linearly extends the wedge horizon). Although the global
algebra of the lightfront is equal to that of ambient Minkowski space and (in
theories with causal propagation) this equality continues to hold between the
wedge algebra and that of its horizon, the \textit{local} structure (the
inclusion-preserving system of local subalgebras) of the holographic
projection brings to light an interesting correlation structure which the net
structure of the wedge algebra does not reveal. Namely the vacuum which (in
absence of external fields) suffers vacuum polarization in all space- time and
light-like direction presents itself with respect to the lightfront degrees of
freedom as a state without transverse vacuum polarization; the omnipresent
field theoretic vacuum polarizations have been compressed into the
(longitudinal) lightray direction in the process of holographic projection in
such a way that the transverse direction remains completely free of vacuum
fluctuations (in the standard lightfront quantization methods these important
but subtle properties are easily overlooked \cite{Suss}). One can then check
that this somewhat unexpected partial return to vacuum polarization-free
quantum mechanics in the transverse direction is physically backed up by the
appearance of transverse Galilei invariance which originates from the
lightfront holography of the translational part of Wigner's ``little group''
\cite{Schroer}. This invariance subgroup of the lightray forms a
two-parametric subgroup of the seven-parametric automorphism group of the
lightfront which arises as the subgroup of the ten-parametric Poincar\'{e}
group, but there are also new geometric invariances (the diffeomorphism of
chiral theories) which are invisible before the holomorphic projection because
they are ``fuzzy'' symmetries of the massive theory in the ambient spacetime structure.

The lightfront holography illustrates in greater detail that the crucial
property, by which localization in local quantum physics leads to thermal
properties, is the appearance of vacuum polarization. Whereas the important
role of virtual and real pair creation was appreciated ever since Hawking's
famous paper on black hole radiation, the vacuum fluctuation-free transverse
tensor structure of the holographic lightfront projection entails an
additional interesting consequence. Namely any extensive physical quantity on
the horizon which behaves additively if the system is in a special uncoupled
state which factorizes under nonoverlapping spatial subdivisions\footnote{The
vacuum of second quantized interaction-free QM is an example for such a state
without entanglement under subdivisions (tensor-product factorization).} will
be characterized by an area- (transverse volume-) density which can be
computed in terms of the longitudinal correlations (which turn out to
correspond to a chiral theory). Here the entropy and its quantum description
in terms of the von Neumann trace formula based on the density matrix of an
impure state immediately comes to ones mind. The problem of
\textit{localization entropy however} requires a bit more abstract setting
which also makes perfect sense in the standard heat bath case. One could view
the quest for entropy as the search for an additive quantity (in the
aforementioned sense) which still would leave open the problem of its
normalization. To solve this, one could refer to the way in which this
quantity enters the fundamental thermodynamic law, i.e. after having derived
the form in which it enters the fundamental laws (which relate it to other
quantities) one could fix the normalization by a convention which of course
should reduce to the classical case if the fundamental law has the classical
form. This is in a way what Bekenstein does when he relates his area
proportional entropy to the quantum Hawking theory by invoking the classical
form of fundamental thermodynamic law. The main change in adapting the
Bekenstein argument to black hole analogs would be on the quantum Hawking side
in that a quantity called the ``surface gravity''\footnote{In the case of the
Unruh Gedanken experiment and black hole analogs this terminology should be
taken with a grain of salt.} which relates dynamical properties of local
quantum physics to the geometric situation (in the more abstract setting of
the next section it links the KMS ``Hamiltonian'' to the modular group) has to
be modified according to the physical nature of the analog. According to the
best of my knowledge there has been no attempt to derive fundamental
thermodynamic laws in case the thermal behavior does not result from an
Gibbs-like quantum ensemble average but rather from vacuum fluctuations caused
by localization.

The next section is a review of the standard heat bath situation and its
relation to the Tomita-Takesaki modular theory and KMS properties
\cite{Haag}\cite{B-B1}\cite{B-B2}.

In the third section we will provide arguments in favor of the existence of an
unnormalized entropy which determines ratios of area densities for different
quantum matter content. Although one does not expect the totally kinematic
situation of Bekenstein, the universality aspects of the holographic
projection only permit a dependence on holographic universality classes. The
holographic projection reduces the problem of area density to the computation
of localization entropy to that in an associated chiral theory.

In order to keep the length of the reference list in balance with the short
size of the presentation, I have opted to refer to reviews and books which
contain a rather extensive bibliography (including the original papers on the
subject) as much as possible. Especially those readers who find these notes
too scarce are encouraged to look at those sources.

\section{Heat bath thermal behavior, KMS property and modular theory}

The correlation functions of a QFT in a quantization box V (in order to obtain
a discrete energy-momentum spectrum) coupled to a heat bath reservoir are
computed with the well-known Gibbs formula
\begin{align}
\omega_{\beta}(A)  &  :=\frac{1}{Z_{V}}tre^{-\beta H_{V}}A,\,\,Z_{V}%
=tre^{-\beta H_{V}}\\
&  \curvearrowright\omega_{\beta}(\mathbf{1)}=1\nonumber
\end{align}
which assigns a (normalized) state\footnote{The existence of inequivalent
representations in the presence of infinitly many degrees of freedom requires
to make a distinction between a state in the sense of assigning an expectation
value to an operator and a a state vector which implements this state if one
goes beyond finite degrees of freedom QM.} on the algebra of bounded
operators.$A\in\mathcal{A}=B(\mathcal{H}).$ The Gibbs formula is meaningful as
long as the partition function $Z$ exists (which requires a discrete
Hamiltonian spectrum bounded below and with finite degeneracy). For this
reason the Gibbs formulation starts from a system enclosed in a box. As in the
vacuum case one may restrict ones interest to case that the algebras are
generated by pointlike fields in which case the operator theory may be
reconstructed from the thermal correlation functions of fields and one has
access to a perturbative formulation. The difference to the vacuum situation
becomes more visible on the level of the operator formalism which may be
obtained from the state $\omega_{\beta}(\cdot)$ on $\mathcal{A}$ ($\simeq$ set
of correlation functions) by the canonical GNS (Gelfand, Neumark and Segal)
construction \cite{Haag} which by now is well known among mathematical
physicist. Using the special property of density matrix states, one may
implement the abstract GNS construction on a Hilbert space $\mathcal{H}_{HS}$
whose vectors are Hilbert-Schmidt operators $\kappa$ i.e. $tr\kappa^{\ast
}\kappa<\infty$%
\begin{align}
\mathcal{H}_{HS}  &  =\left\{  \psi_{\kappa}\text{ }|\text{ }(\psi_{\kappa
},\psi_{\kappa^{\prime}})\equiv tr\kappa^{\ast}\kappa^{\prime}\right\}
\label{left}\\
&  \pi(A)\psi_{\kappa}\equiv\psi_{A\kappa}\in\mathcal{H}_{HS}\nonumber
\end{align}
where $\pi(\cdot)$ denotes the representation of the algebra on $\mathcal{H}%
_{HS}.$ The HS Hilbert space is isomorphic to the tensor product of the
original Hilbert space $\mathcal{H}_{HS}\simeq\mathcal{H}\otimes\mathcal{H}$
since the linear combinations of ``dyads'' $\left|  \psi\right\rangle
\left\langle \varphi\right|  $ from the tensor product upon closure in
$\mathcal{H}_{HS}$ generate the HS Hilbert space. This ``doubling'' entails
that besides the left action (\ref{left}) of the full algebra of bounded
operators $B(\mathcal{H})$ on $\mathcal{H}_{HS}$ there is a right action which
in the HS description reads $\psi_{\kappa}\rightarrow\psi_{\kappa A}.$ In
order to distinguish between the left and right representation and to maintain
the naturality of composition (representation) laws, one defines the right
representation as a conjugate (antilinear) representation
\begin{align}
\pi_{l}(A)\psi_{\kappa}  &  =\pi(A)\psi_{\kappa}=\psi_{A\kappa}%
\label{doubling}\\
\pi_{r}(A)\psi_{\kappa}  &  =\psi_{\kappa A^{\ast}}\nonumber
\end{align}
It is obvious that any right action commutes with any left action i.e.
$\pi_{r}(\mathcal{A})\subseteq\pi(\mathcal{A})^{\prime}$ (where the dash
denotes the von Neumann commutant of $\pi(\mathcal{A})$ in $\mathcal{H}_{HS})$
and it is not difficult to see that in fact equality holds \cite{Bra-Ro}. In
$\mathcal{H}_{HS}\simeq\mathcal{H}\otimes\mathcal{H}$ there are many more
operators than in $\pi(\mathcal{A}),$ e.g. the anti-unitary ``flip''
\begin{align}
J\psi_{\kappa}  &  :=\psi_{\kappa\ast},\text{ }J^{2}=1\\
&  J\pi(A)J=\pi_{r}(A)\nonumber
\end{align}
which in the tensor product representation would simply interchange the bra
and ket in a dyad.

Using now the fact that the Hilbert-Schmidt operator $\kappa_{0}\equiv$
$\rho^{\frac{1}{2}}$associated with the nondegenerate (no zero eigenvalue)
Gibbs density matrix
\begin{equation}
\omega_{\beta}(A)=\left(  \psi_{\kappa_{0}},\pi(A)\psi_{\kappa_{0}}\right)
=tr\kappa_{0}A\kappa_{0}%
\end{equation}
is cyclic and separating with respect to the action $\pi(\mathcal{A})$ of the
algebra (i.e. sufficiently entangled in $\mathcal{H}\otimes\mathcal{H}$ so
that the application of this subalgebra permits to approximate any vector in
$\mathcal{H}\otimes\mathcal{H}$ and that it is not possible to annihilate the
entangled state with a nonzero operator from $\pi(\mathcal{A})$), one checks
the validity of the relation (mainly an exercise in the correct application of
definitions)
\begin{align}
S\pi(A)\psi_{\kappa_{0}}  &  =\pi(A)^{\ast}\psi_{\kappa_{0}},\,A\in
\mathcal{A}\\
where\,\,S  &  :=J\pi(\rho^{\frac{1}{2}})\pi_{r}(\rho^{-\frac{1}{2}%
})\curvearrowright S^{2}\subset1\nonumber
\end{align}
where the last relation is a notation for the fact that the
unbounded\footnote{Since $\pi_{r}(\rho^{-\frac{1}{2}})$ is unbounded even for
Hamiltonians with one-sided unbounded spectrum.} operator $S$\thinspace\ is
involutive on its domain. By an additional notational convention one gets this
relation into the form where it may be used as a special illustration of a
very general operator algebra situation
\begin{align}
&  H_{\operatorname{mod}}\equiv\pi(H)-\pi_{r}(H)\label{mod}\\
&  \Delta^{it}\equiv e^{-i\beta tH_{\operatorname{mod}}},\,\,S=J\Delta
^{\frac{1}{2}}\nonumber\\
&  H_{\operatorname{mod}}\psi_{\kappa_{0}}=0,\,\,\Delta^{it}\psi_{\kappa_{0}%
}=\psi_{\kappa_{0}}\nonumber\\
&  \Delta^{-it}\pi(A)\Delta^{it}=\pi(\alpha_{\beta t}(A)),\,\,\alpha
_{t}(A)=Ade^{iHt}(A)\equiv e^{iHt}Ae^{-iHt}\nonumber
\end{align}
Whereas in vacuum QFT the energy operator $H$ (obtained by integrating the
energy density) is the same as the generator of the translation, in the heat
bath situation it is the doubled Hamiltonian $H_{\operatorname{mod}}$ which
leaves the thermal reference state invariant and generates the translation
symmetry of the thermal correlation functions (whereas the energy operator
fluctuates in the thermal state and these fluctuations become infinitely big
in the thermodynamic limit V$\rightarrow\infty)$. The unitary $e^{itH}\,$which
represents the time translation in the vacuum representation continues to
implement the time automorphism through its adjoint action on the algebra but
ceases to be a well defined unitary operator in the GNS Hilbert space
associated with the thermodynamic limit where its role is taken over by
$e^{itH_{\operatorname{mod}}}$ (which acts on both $\pi(\mathcal{A})$ and
$\pi_{r}(\mathcal{A})=\pi(\mathcal{A})).$The following KMS relation follows
from the cyclicity property of the trace in the Gibbs formula
\begin{align}
&  \omega_{\beta}(\alpha_{t}(A)B)=\omega_{\beta}(B\alpha_{t+i\beta}(A))\\
\exists\,F_{A,B}(z)\,,\,F_{A,B}(t)  &  =\omega_{\beta}(B\alpha_{t}%
(A)),\,F_{A,B}(t+i\beta)=\omega_{\beta}(\alpha_{t}(A)B)\nonumber
\end{align}
where the second line expresses the analytic content of the KMS condition in
more careful terms: there exist an analytic functions $F_{A,B}(z)$ which
interpolates between the thermal expectation values of operator products taken
in different orders ; this function is analytic in the strip $0<Imz<\beta$ and
has continuous boundary values on both margins which relate to the two
different orders according to the above formulas.

There are good reasons in favor of a formulation which deals with infinitely
extended system from the very outset (and not as a result of a thermodynamic
limiting process on box-enclosed systems) and to consider finite systems as
open subsystems \cite{Haag}. These formulations substitute the Gibbs formula
(the Gibbs density matrix as a trace class operator on Fock space is only
meaningful for box-enclosed systems) by the KMS property, which does not refer
to particular representations and trace class operators. In this more general
setting heat bath thermal physics becomes incorporated into a mathematical
very deep and physically extremely useful theory: the Tomita-Takesaki modular
theory. At the heart of that theory is the following theorem \cite{Bra-Ro}.

\begin{theorem}
(main theorem of the Tomita-Takesaki modular theory) Let ($\mathcal{A,H}%
,\Omega$) denote a weakly closed (von Neumann) operator algebra $\mathcal{A}$
acting on a Hilbert space $\mathcal{H,}$ with $\Omega\in\mathcal{H}$ a vector
on which $\mathcal{A}$ acts in a cyclic and separating manner. Then there
exists an antilinear closed involutive operator $S$ which has the dense
subspace $\mathcal{A}\Omega$ in its domain such that
\begin{align}
SA\Omega &  =A^{\ast}\Omega,\,A\in\mathcal{A}\\
S  &  =J\Delta^{\frac{1}{2}},\,J\Delta=\Delta^{-1}J\nonumber
\end{align}
The polar decomposition of $S$ leads to an antiunitary $J$ and a positive
$\Delta$ which in turn defines the unitary modular group $\Delta^{it}.$ Their
significance results from their adjoint action on the algebra
\begin{align}
J\mathcal{A}J  &  =\mathcal{A}^{\prime}\\
\sigma_{t}(A)  &  \equiv\Delta^{it}A\Delta^{-it}\in\mathcal{A}\nonumber
\end{align}
The modular automorphism $\sigma_{t}$ fulfills the following KMS property
(with $\beta=-1)$%
\begin{equation}
\omega(\sigma_{t}(A)B)=\omega(B\sigma_{t-i}(A)B),\,\omega(\cdot)\equiv\left(
\Omega,\cdot\Omega\right)
\end{equation}
and depends only on the state $\omega$ (and not on its implementing vector
$\Omega).$
\end{theorem}

Any faithful state $\omega$ on an operator algebra leads to this situation via
the GNS construction, and any KMS state is faithful \cite{Bra-Ro}. Even though
$J$ implements an antiisomorphism of $\mathcal{A}$ with its commutant
$\mathcal{A}^{\prime}$ and both together generate the full ambient algebra
$B(\mathcal{H})=\mathcal{A}\vee\mathcal{A}^{\prime},$ the validity of a tensor
product representation $B(\mathcal{H})=\mathcal{A}\otimes\mathcal{A}^{\prime}$
(with a factorizing modular unitary) is restricted to the case of
$\mathcal{A}$ being a type I algebra and $\omega(\cdot)$ is induced by a
density matrix ( i.e. the KMS property leads back to a Gibbs density matrix).
The speciality of the above Gibbs density matrix situation as compared to the
general KMS property in the thermodynamic limit is precisely this tensor
product structure (which gets lost in the limit when the algebra changes its
type to type III$_{1}$). This can be seen explicitly by solving the KMS
relation which (for a scalar field) leads to the well known thermal two point
function (the higher point functions have the standard free field product
structure in terms of the two point function)
\begin{align}
&  \omega_{\beta}\left(  A(x)A(y)\right)  =\frac{1}{\left(  2\pi\right)  ^{3}%
}\int(e^{-ip(x-y)}\frac{1}{1-e^{-\beta\omega}}+e^{ip(x-y)}\frac{e^{-\beta
\omega}}{1-e^{-\beta\omega}})\frac{d^{3}p}{2\omega}\,\label{free}\\
&  \pi_{\beta}(A(x))=\frac{1}{\left(  2\pi\right)  ^{\frac{3}{2}}}%
\int(e^{-ipx}\pi_{\beta}(a(p))+e^{ipx}\pi_{\beta}(a(p))^{\ast})\frac{d^{3}%
p}{2\omega},\,\,\,\,p_{0}=\omega=\sqrt{\vec{p}^{2}+m^{2}}\nonumber\\
&  \left(
\begin{array}
[c]{c}%
\pi_{\beta}(a(p))\\
\pi_{\beta}(a(p))^{\ast}%
\end{array}
\right)  \equiv\left(
\begin{array}
[c]{cc}%
\cosh\alpha(p) & \sinh\alpha(p)\\
\sinh\alpha(p) & \cosh\alpha(p)
\end{array}
\right)  \left(
\begin{array}
[c]{c}%
a(p)\otimes1\\
1\otimes a^{\ast}(p)
\end{array}
\right)  ,\,\cosh\alpha(p)=\sqrt{\frac{1}{1-e^{-\beta\omega}}}\nonumber\\
&  \omega_{\beta}\left(  A(x)A(y)\right)  =\omega_{0,0}(\pi_{\beta}%
(A(x))\pi_{\beta}(A(y)))\nonumber
\end{align}
where the third line is Bogoliubov transformation on ordinary
creation/annihilation operators in a tensor doubled Fock space and the last
line states that the thermal two point function can be written as a vacuum
representation of the thermal $\pi_{\beta}(A(x))$ operators in a tensor
product vacuum (``purification by doubling''). Contrary to the finite volume
Gibbs situation the above infinite volume Bogoliubov transformation has no
unitary implementation which could undue the entanglement with respect to the
doubling associated with the Bogoliubov transformation; in fact these
persistent thermal fluctuations in unbounded spacetime are the cause for the
loss of the tensor product structure (and the change from type I to type III)
between the global von Neumann algebra generated by the fields $\mathcal{A}$
and its commutant $\mathcal{A}^{\prime}$ in the thermal Hilbert
space\footnote{Whereas the GNS construction on the finite volume (Gibbs)
theory leads to a quantum mechanical type I$_{\infty}$ von Neumann algebra
whose characteristic property is the existence of minimal projectors ($\simeq$
maximal measurements), the infinite volume finite temperature QFTs (including
(\ref{free})) are typically of ``hyperfinite type III$_{1}".$ As will be shown
in the next section, localization-caused vacuum fluctuations without a heat
bath environment have similar thermal consequences.}. Not only are the thermal
infinite volume algebras inequivalent to vacuum algebras, but even the
algebras for different temperatures turn out to be inequivalent \cite{Bra-Ro}.
Although ``purification'' of impure states can always be achieved by
enlargement, there is no Bogoliubov transformation formula for interacting
infinite thermal systems.

A KMS-based direct approach reveals the following additional interesting
structural differences to vacuum QFT.

\begin{itemize}
\item  Unlike vacuum QFT the analytic properties of spacetime or momentum
space correlation functions cannot be described in terms of different boundary
values one holomorphic ``master function''; this complicates in particular the
relation between the imaginary time (Matsubara formalism) and real time
correlation functions. Even though Lorentz invariance is spontaneously broken,
the continued validity of relativistic causality leads to an enlargement of
KMS analyticity which was only recently noticed (the relativistic KMS property
in \cite{B-B2}).

\item  Contrary to the vacuum theory where (at least in the case of mass-gaps)
the LSZ scattering theory allows a description in the Fock space of the
asymptotic particles without encountering infrared divergencies, real time
infinite volume thermal theories calculated perturbatively starting from free
thermal propagators lead to perturbative infrared problems. Whereas
ultraviolet problems in renormalizable theories tend to be of technical origin
(cavalier use of singular pointlike fields), difficulties in the infrared
invariably point to an insufficient physical understanding. This problem was
recently addressed by a concrete proposal on the asymptotic behavior
\cite{B-B3} which takes into account the fact that some interaction dependent
dissipative effects are asymptotically persistent. This proposal if used in
perturbation theory, should avoid infrared problems; a task which still has to
be carried out.
\end{itemize}

Most perturbative calculations have been done in the imaginary time (ITF)
Matsubara formalism which by functional formulation relates to the Gibbs
setting. In that case (as for spacelike separations), the KMS property
simplifies to a periodicity relation which in turn leads to discrete energies.
The rather involved relation of ITF with areal time (RTF) formalism (which
requires the application of Carlsonian theorems in order to reconstruct the
correlations for arbitrary energies from the discrete ITF values) was only
recently clarified \cite{Viano}. For two decades a special formulation of RTF
known under the name ``Thermo Field Theory'' \cite{Umezawa} gained increasing
popularity with practitioners since it allows to do calculations in terms of
generalized (``doubled'') Feynman rules with the same efficiency as in vacuum
problems. It results from the tensor product doubling of Fock space before the
thermodynamic limit; the ``Tilde'' copy of an observable in the second tensor
factor is simply the Tomita $J$-mirrored operator (for a more detailes about
its connection with the older KMS formalism see \cite{Ojima}). Modular theory
shows that the tilde fields survive the thermodynamic limit but loose their
Fock space tensor product structure relative to the standard fields.
Mathematical physicist usually prefer the older KMS formulation \cite{Haag}
since its relates in a stronger and conceptually less mystifying way to
physical and mathematical principles. Our preference of KMS in these notes is
a result of the fact that localization-caused thermal properties do not arise
as thermodynamic limits of Gibbs situations; rather the situation is the other
way around in that one has to invent such tensor product interpolations (the
``split property'' of next section) in order to be able to envisage a
localization entropy corresponding to its KMS temperature.

In the next section it will be shown that the modular framework is capable of
incorporating both the heat bath- and the localization- caused thermal properties.

\section{Thermal aspects caused by vacuum polarization on horizons}

As mentioned in the introduction, the Hawking black hole thermal aspects are
not limited to event horizons generated by spacetime curvature in general
relativity, but they occur in all systems (the black hole analogs) of local
quantum physics i.e. quantum systems with a notion of causal propagation of
quantum matter (and the ensuing inevitable vacuum polarization). In a way this
kind of thermalization without invoking heat bath ensembles is related to the
loss of information by restricting states to subalgebras. But there is a
caveat, according to the Tomita-Takesaki theorem the thermalization (i.e. the
KMS property of the subalgebra-restricted state) takes place iff this
restriction causes the state vector to be cyclic and separating. Although this
does not happen in QM (where the restriction to a compact region would lead to
an inside/outside tensor product factorization without any entanglement of the
vacuum) in line with the absence of causal propagation with a maximal
velocity, this is a vastly general phenomenon in QFT thanks to the
omnipresence of vacuum polarization. The Reeh-Schlieder theorem of QFT (the
localization generated ``operator-state'' relation in the more folkloristic
terminology often used in conformal QFT) insures that these properties are
always fulfilled as long as the localization region has a nontrivial causal
disjoint. Although the Tomita-Takesaki theorem asserts that the modular KMS
automorphism always exists in these cases, the physical interpretation up to
now has been restricted to cases of geometric (diffeomorphism, non-fuzzy)
action of the modular group $\sigma_{t}$ which in the context of massive QFT
in Minkowski space leaves only the Lorentz boosts of wedge region (and in CST
the much richer cases of horizons associated with Killing symmetries). The
best known illustration without curvature is Unruh's Gedankenexperiment
already mentioned in the introduction in which the observables localized in a
wedge region (the Bisognano-Wichmann of a ) bounded by a causal horizon are
realized by a family of uniformly accelerated observers whose Hamiltonian is
proportional to the Lorentz boost (for a simple but enlightening presentation
using modular concepts see \cite{Sewell}).

This raises the question whether the area behavior of entropy (observed first
with the help of classical arguments applied to black hole horizons by
Bekenstein) could also be expected to be the manifestation of the same vastly
general mechanism which is responsible for the appearance of a temperature via
quantum localization. Trying to answer this question with standard box
quantization methods and ad hoc cut-offs (to obtain a finite entropy via
standard degree of freedom counting) proved to be inconclusive \cite{Sorkin}.
According to the above ideas the relevant question should be whether by
physically motivated ideas (i.e. by remaining within the given local theory
\ and thus avoiding ad hoc cut-offs) one can associate a localization entropy
with the Lorentz boost in its role as the modular group of the wedge algebra
so that the Hawking temperature (which cannot be used to get informations
about quantum gravity) and the area density of entropy live under a common
roof. Since the upper horizon algebra turns out to be globally identical to
the wedge algebra (but has a very different local structure!), this question
can be reformulated in terms of the scale transformation to which the boost
reduces on the horizon (half the lightfront). The formalism which achieves
this reformulation including the determination of the new local structure on
the horizon is ``algebraic lightfront holography'', a mathematically and
conceptually rigorous variant of the old lightfront (or p$\rightarrow\infty$
frame) formalism in which greater attention is payed to the locality of both
the original ambient theory and its projection as well as to the fine points
of their only partially local mutual relation (The lightfront is not a
globally and even not a locally hyperbolic manifold). It turns out that the
lightfront holography leads to a QFT with a seven parametric symmetry subgroup
of the Poincar\'{e} group which contains in particular a transverse Galilei
group which results from the holographic projection of the ``translations''
contained in Wigner's 3-parametric ``little group'' of the lightray in the
lightfront \cite{Schroer}. Limiting the presentation for reasons of
pedagogical brevity to free fields\footnote{For interacting fields with
infinite wave function renormalization constants the holographic process
requires a conversion into operator algebras; but even though the methods turn
out to be quite different, the results on the absence of transverse
fluctuations are the same \cite{Schroer}.} where the holographic projection
can be obtained in terms of lightfront restriction of fields without first
converting them into algebras, one finds for the Weyl generators associated
with a scalar free field $A(x)$%
\begin{align}
&  W(f):=e^{iA(f)},\,\,A(f)=\int A(x)f(x)d^{4}x\\
&  W(g,f_{\perp})\longrightarrow W_{LF}(g,f_{\perp})=e^{iA_{LF}(g,f_{\perp}%
)},\,\,A_{LF}(g,f_{\perp})=\int a^{\ast}(p_{-},p_{\perp})g(p_{-})f_{\perp
}(x_{\perp})\frac{dp_{-}}{2\left|  p_{-}\right|  }d^{2}p_{\perp}%
+h.c.\,\nonumber\\
&  \curvearrowright\left\langle W(g,f_{\perp})W(g^{\prime},f_{\perp}^{\prime
})\right\rangle =\left\langle W(g,f_{\perp})\right\rangle \left\langle
W(g^{\prime},f_{\perp}^{\prime})\right\rangle \text{ }if\,\,suppf\cap
suppf^{\prime}=\emptyset\nonumber
\end{align}
The second line formulates lightfront restriction on the dense set of wedge
supported test functions which factorize into a longitudinal and a transverse
part \cite{Schroer}. The third line is the statement that holographically
projected Weyl generators (and therefore also the algebras they generate) have
no transverse fluctuations; the holographic projection compresses all vacuum
fluctuations into the lightlike direction. This reduces the problem of
horizon-associated entropy to the problem of looking for an area (the area of
the edge of the wedge which limits the upper horizon) density of entropy and
should be interpreted as the localization entropy of a halfline in a chiral
theory\footnote{I learned from Detlev Buchholz that the problem of
localization entropy in a more general setting as the present one had been
considered as a very hard problem by algebraic field theorist (see e.g.
\cite{Narn}). The present holographic setting is expected to make it more
amanable.}. From the time of discovery of infinite vacuum fluctuations on the
boundary of sharply localized partial Noether charges by Heisenberg it became
clear that the remedy for obtaining well defined partial charge operators is
to make the spacetime boundary somewhat ``fuzzy'' by using smearing functions
which smoothly interpolate between one and zero. For the more abstract problem
of assigning an entropy to a ``partial vacuum'' (i.e. the ambient vacuum
reduced to the horizon algebra) the analog construction is based on the split
property namely the assertion that between two sharply localized (necessarily
hyperfinite type III$_{1})$ algebras (where the larger one $\mathcal{A}%
_{\varepsilon}$ is obtained by extending the smaller by an arbitrarily small
``$\varepsilon$-collar'') there exist tensor-factorizing type I algebras
$\mathcal{N}_{\varepsilon}$ on which the vacuum state becomes a faithful
density matrix $\rho_{\varepsilon}$
\begin{align}
\mathcal{A}  &  \subset\mathcal{N}_{\varepsilon}\subset\mathcal{A}%
_{\varepsilon},\,\ B(\mathcal{H})=\mathcal{N}_{\varepsilon}\otimes
\mathcal{N}_{\varepsilon}^{\prime}\\
\omega(N)  &  =tr(\rho_{\varepsilon}N),\,\,N\in\mathcal{N}_{\varepsilon
}\nonumber
\end{align}
This would bring us back to the Gibbs setting, except that in this case one
has no explicit formula for $\rho_{\varepsilon}$ in terms of a Hamiltonian;
the existence of $\mathcal{N}_{\varepsilon}$ and the form of $\rho
_{\varepsilon}$ is rather determined in terms of more abstract modular
concepts. The problem of assigning an entropy to the trace class operator
$\rho_{e}$ is not different from that in the heat bath situation; in addition
to the trace class property of the density one has to make sure that the
$\rho_{\varepsilon}$ operator does not have too many eigenvalues around zero
which would prevent the convergence in the von Neumann formula $S(\rho
_{\varepsilon})=-tr\rho_{\varepsilon}\ln\rho_{\varepsilon}.$

The transverse symmetry and absense of correlations for the case at hand
reduces the area density problem to that of a chiral theory on a lightray,
which by a Cayley transform becomes a chiral theory on the circle. Cutting the
circle into the two halfs $I$ and its opposite $I^{\prime}$ and separating the
ends by a distance $\varepsilon$ so that $I_{\varepsilon}\subset I,I^{\prime
}\subset I_{\varepsilon}^{\prime}$ our problem is to compute a ``split
entropy'' $S(\rho_{\varepsilon})$ associated with a chiral split inclusion
$\mathcal{A}(I_{\varepsilon})\subset\mathcal{N}_{\varepsilon}\subset
\mathcal{A}(I),\;\mathcal{B}(\mathcal{H})=\mathcal{N}_{\varepsilon}%
\otimes\mathcal{N}_{\varepsilon}^{\prime}.$ The type $I_{\infty}$ algebra
$\mathcal{N}_{\varepsilon}$ approaches for $\varepsilon\rightarrow0$ the
sharply localized halfline algebra $\mathcal{N}_{\varepsilon}\rightarrow
\mathcal{A}(I)$ so that the modular group corresponding to $\rho_{\varepsilon
}$ approximates the dilatation group which in turn is the holographic
lightfront projection of the wedge-affiliated boost. The entropy diverges for
$\varepsilon\rightarrow0$ and it is expected that it behaves universally as
$\simeq-\ln\varepsilon.$ Since we were not able to supply a proof , we
formulate the following universality behavior as a conjecture.

\begin{conjecture}
(universality conjecture) The split localization entropy of the halfline in
the sharp localization limit goes as
\begin{equation}
S_{\varepsilon}\underset{\varepsilon\rightarrow0}{=}-c\ln\varepsilon
\end{equation}
where c depends on the chiral model appearing in the holographic projection
(the holographic universality class). Hence the area density of entropy
associated with the horizon is only determined by the above considerations up
to an overall normalization which is independent on the holographic
universality class.
\end{conjecture}

The conjecture is consistent with the observation that the there is no
specific reference to gravitational aspects, i.e. the above considerations
hold also for the black hole analogs.

The conceptually most poignant way to understand the role of this localization
entropy is to associate it with a measure of entanglement of the physical
vacuum $\omega$ of the ambient theory relative to the product vacuum
$\omega_{P}$ which expresses the division of the lightfront into the wedge
horizon and its opposite. After taking the transverse symmetry and absence of
correlation into account, the problem becomes one of studying a product state
on the two-fold localized chiral algebra $\mathcal{A(}I_{\varepsilon
}\mathcal{)\vee}$ $\mathcal{A(}I\mathcal{)}^{\prime}$%

\begin{equation}
\omega_{p}(AA^{\prime})=\omega(A)\omega(A^{\prime}),\,\,A\in\mathcal{A(}%
I_{\varepsilon}\mathcal{)},A^{\prime}\in\mathcal{A(}I\mathcal{)}^{\prime}%
\end{equation}
In this formulation (using states instead of implementing vectors) there is no
direct dependence on the (arbitrarily chosen) intermediate type $I_{\infty}$
factor\footnote{However the normality of the product state on the two-fold
localized algebra is equivalent to the existence of an intermediate type
$I_{\infty}$ factor \cite{Haag}.} and the question arises whether it is
possible to define a kind of relative entanglement of $\omega$ with respect to
$\omega_{P}$ on $\mathcal{A(}I_{\varepsilon}\mathcal{)\vee A(}I\mathcal{)}%
^{\prime}$ ($\simeq\mathcal{A(}I_{\varepsilon}\mathcal{)\otimes A(}%
I\mathcal{)}^{\prime})$ directly in terms of these states without implementing
them by vectors (entanglement leads directly to impurity upon restriction of .
There exists a variational entropy formula by Kosaki, but unfortunately it was
not possible to put it to good use for this kind of problem \cite{Narn}. There
is however special implementation of the split isomorphism $\Phi
(\mathcal{A(}I_{\varepsilon}\mathcal{)\vee A(}I\mathcal{)}^{\prime
})=\mathcal{A(}I_{\varepsilon}\mathcal{)\otimes A(}I\mathcal{)}^{\prime}$ in a
rather large space in which (at least for interaction-free theories) overlaps
of states can be bounded above by absolute values of overlap of states. The
calculation involves the application of the so-called flip-trick which
resembles the Noether formalism for free currents \cite{charges}. The result
is that the overlap for an algebra generated by a chiral scalar free field can
be written as \cite{Schroer}%

\begin{equation}
e^{-\frac{1}{2}\left\langle j(f),j(f)\right\rangle _{0}}\simeq\varepsilon
\end{equation}
Here $j$ is a abelian chiral current and $f$ is a smearing function which is
one on $I_{\varepsilon},$ zero on $I$ passes smoothly from one to zero in the
two $\varepsilon$ intervals between. The only contribution comes from the
$\varepsilon$ regions where the function changes. Choosing a fixed smooth
function and doing the limit by scaling the transition region, one finds a
vanishing overlap going with the first power in $\varepsilon$. One expects a
power type of vanishing in every chiral model. In principle one can compute
the entanglement of the vacuum in a suitable tensor product basis (e.g. one
defined by applying polynomials of the Fourier coefficients of $j$ to the
tensor product vacuum). One would expect that the strength of the entropy is
universally logarithmic, but that the numerical factors multiplying
$-\ln\varepsilon$ depend on the chiral matter which represents the holographic
universality classes of the original quantum matter. The universality
conjecture is related to the observation that the vacuum polarization related
to the split horizon (which is concentrated to the collar of size$\varepsilon
)$ is not a property of short distance singularities of particular field
coordinatizations\footnote{The standard ultraviolet divergencies are caused by
the use of (inevitably) singular pointlike field coordinatizations of the
local net of algebras.
\par
{}} but rather an intrinsic property of the split-localized algebra itself.
Actually it does not appear difficult to use the same implementation as before
in order to obtain an upper bound for the $\varepsilon$-dependent split
entropy since a natural discrete basis in the tensor product space can be
given in terms of the rotational modes of the chiral free field. We hope to
return to this problem.  

The missing overall normalization should distinguish analog systems from
gravitational black holes (there is no distinction in the above
considerations) and this is expected to come about through thermodynamic
fundamental laws. It is an open question whether and how such laws can be
derived for thermal aspects which are not resulting from the standard heat
bath ensemble averaging. One possibility would be that such fundamental
thermodynamic laws (through which entropy is related to other quantities
computed from local quantum physics) need nonvanishing curvature and compact
regions of bifurcations (edges) of horizons, but up to now there seems to be
no support for this idea.

Whereas the Bekenstein area law is totally independent of the kind of quantum
matter (but depends on the relation of the geometrical modular group to the
dynamical time development group which distinguished gravitational situations
from black hole analogs), its quantum version may have a holographic
universality class dependence. In this sense it may contain slightly more
structure about the quantum system than the Hawking temperature of the KMS state.

\section{Concluding remarks}

In this paper I argued that the modular approach allows to unify the
understanding of heat bath thermal aspects of quantum physics with those
caused by vacuum fluctuations which are the inexorable result of causal
localization in QFT (even when the global state is the vacuum).

The problem of the area density of quantum entropy associated with an horizon
(without or with curvature) has been in the center of interest for a long
time. Ever since Bekenstein proposed his famous area law invoking analogies of
black hole physics with classical thermodynamics, many physicists expected
that its quantum understanding contains important clues about QG beyond QFT in
CST. The above argument in favor of vacuum fluctuation caused by localization
calls for caution about the various speculative ideas which propose to solve
this problem by invoking ``new physics''. Presently a more conservative
explanation of the area behavior (on the level of the Hawking-Unruh thermal
aspects) which maintains all the principles of QFT but adds new conceptual and
mathematical tools cannot be excluded. In the previous section it was argued
that the solution of the entropy problem in the setting of QFT depends on a
certain universality behavior in the holographic projection as well as the
validity of normalizing quantum fundamental thermodynamic laws for
localization-induced thermal behavior, both unproven. In contrast to the many
speculative attempts\footnote{It would be nice if instead of many computations
on this subject one could find some interpretation. Since physics should
de-mystify nature, one would like to know: if it is not vacuum polarization
(which is behind the Hawking thermalization) due to localization on horizons,
then what else is responsible for the area behavior of Bekenstein entropy?}, a
clarification along these more conservative lines has the additional advantage
that its importance would not be diminished by a negative outcome i.e. a
clear-cut argument why the conjectured behavior may be incompatible with the
principles underlying QFT would also constitute a major achievement.

A shorter report based on a talk with a similar content can be found under hep-th/0301082

Acknowledgments: I am indebted to Detlev Buchholz for explaining some fine
points of his work with Jaques Bros on the correct analog of the LSZ
asymptotics in the heat bath setting of local quantum physics.

\end{document}